\documentclass[final,1p,times]{elsarticle} 
\usepackage{graphicx} 
\usepackage{amssymb} 
\usepackage{amsthm} 
\usepackage{lineno} 
\usepackage{graphicx}

\def\eq#1{{(\ref{#1})}}

\newcommand{\beq}{\begin{equation}}
\newcommand{\eeq}{\end{equation}}
\newcommand{\ben}{\begin{eqnarray*}}
\newcommand{\een}{\end{eqnarray*}}
\def\dm{{\partial}_{\mu}}
\def\eps{\epsilon^{\mu \nu \lambda \sigma}}

\def\d{\partial}
\journal{Nuclear Physics A} 
\begin{document} 

\begin{frontmatter} 


\title{Chern-Simons current and local parity violation \\ in hot QCD matter}

\author{Dmitri E. Kharzeev$^{a,b}$}

\address[a]{Nuclear Theory Group \\
Physics Department, 
Brookhaven  National Laboratory\\
Upton, New York 11973, USA\\[0.2ex]} 
\vskip0.2cm
\address[b]{Physics Department, 
Yale University\\
 New Haven, CT 06520-8120, USA}

\begin{abstract} 
Non-Abelian gauge theories "live" in a space-time with non-trivial topology that 
can be characterized by an odd-dimensional Chern-Simons form. In QCD, Chern-Simons form is induced by the chiral anomaly and the presence of topological solutions; it opens a possibility for the breaking of P and CP invariances in strong interactions ("the strong CP problem").  
While there is apparently no global P and CP violation in QCD, here I argue that 
 topological fluctuations in hot quark-gluon matter can become directly observable in the 
presence of a very intense external magnetic field by inducing {\it local} P- and CP- odd effects. 
These phenomena can be described by using the Maxwell-Chern-Simons electrodynamics as an effective theory. Local P and CP violation in hot QCD matter can be observed in experiment through the "chiral magnetic effect" -- the separation of electric charge along the axis of magnetic field that is created by the colliding relativistic ions. There is a recent evidence for the electric charge separation relative to the reaction plane of heavy ion collisions from the STAR Collaboration at RHIC.

\end{abstract} 

\end{frontmatter} 



\section{Introduction}

Topological fluctuations are believed to play an important role in the structure of QCD vacuum and in the properties of hadrons. They also open the possibility of  P and CP violation in QCD ("the strong CP problem")\footnote{This possibility is probably not realized in the present-day Universe -- the experimental bounds from the electric dipole moments on the amount of P and CP violation in QCD  are very stringent.}. However until now all of the evidence for the topological effects in QCD from experiment, however convincing, has been indirect. Here I will present the arguments \cite{Kharzeev:2004ey,Kharzeev:2007tn,Kharzeev:2007jp,Fukushima:2008xe,Kharzeev:2009pj} for the possibility to observe the topological effects in QCD directly in the presence of very intense external electromagnetic fields due to the "chiral magnetic effect".  Recently the chiral magnetic effect has been observed on the lattice \cite{Buividovich:2009wi}. The electromagnetic fields of the required  strength can be created in heavy ion collisions \cite{Kharzeev:2007jp,Skokov:2009qp}. At this Conference, an evidence 
for the $P$-odd charge separation effect has been presented by the STAR Collaboration at RHIC \cite{Voloshin:2009hr}.

\section{Topological effects in ${\rm\bf QCD} \times {\rm\bf QED}$: Maxwell-Chern-Simons theory}

Consider QCD coupled to electromagnetism; the resulting theory possesses $SU(3) \times U(1)$ gauge symmetry.
Let us discuss the electromagnetic sector of this theory at large distances. Electromagnetic fields will couple to the electric currents $J_\mu = \sum_f  q_f \bar{\psi}_f \gamma_\mu \psi_f$, where $q_f$ are the electric charges of the quarks.
In addition, the $\theta$-term in the QCD Lagrangian through the quark loop will induce  the coupling of electromagnetic fields to the topological charge carried by the gluon fields. We will introduce an effective pseudo-scalar field $\theta = \theta(\vec x, t)$ to describe the topological charge distribution and write down the resulting effective Lagrangian as
\beq\label{MCS}
{\cal L}_{\rm MCS} = - \frac{1}{4}F^{\mu\nu}F_{\mu\nu} - A_\mu J^\mu - \frac{c}{4}\ \theta\  \tilde{F^{\mu\nu}}F_{\mu\nu},
\eeq
where 
$c = \sum_f q_f^2 e^2 / (2\pi^2)$, and $A_\mu$ and $F_{\mu\nu}$ are the electromagnetic potential and field strength tensor, respectively. 
This is the Lagrangian of Maxwell-Chern-Simons, or axion, electrodynamics in $(3+1)$ dimensions that has been discussed previously in \cite{Wilczek:1987mv,Carroll:1989vb,Sikivie:1984yz}. 
If $\theta$ is a constant, then the last term in \eq{MCS} represents a full divergence 
$
\tilde{F^{\mu\nu}} F_{\mu\nu} = \partial_\mu J_{CS}^\mu
$
of the Chern-Simons current 
\beq\label{topdiv1}
J_{CS}^{\mu} = \epsilon^{\mu\nu\rho\sigma} A_{\nu} F_{\rho\sigma};
\eeq
note that because of the presence of the anti-symmetric tensor in \eq{topdiv1}, the dynamics of a particular component of the Chern-Simons current is effectively three-dimensional. 
Being a full divergence, this term 
does not affect the equations of motion.

The situation is different if the field $\theta = \theta(\vec x, t)$ varies in space-time.      
Indeed, in this case omitting a full derivative and  introducing notation
$
P_\mu = \partial_\mu \theta = ( \dot{\theta}, \vec P )
$
we can re-write the Lagrangian \eq{MCS} in the following form:
\beq\label{CS}
 {\cal L}_{\rm MCS} = - \frac{1}{4}F^{\mu\nu}F_{\mu\nu} - A_\mu J^\mu + \frac{c}{4} \ P_\mu J^\mu_{CS}.
\eeq
Since $\theta$ is a pseudo-scalar field, $P_\mu$ is a pseudo-vector; as is clear from   \eq{CS}, 
it plays a role of the potential coupling to the Chern-Simons current \eq{topdiv1}. However, unlike the vector potential $A_\mu$, $P_\mu$ is not a dynamical variable and is a pseudo-vector that is fixed by the dynamics of chiral charge -- in our case, determined by the fluctuations of topological charge in QCD.
   
Let us write down the Euler-Lagrange equations of motion that follow from the Lagrangian \eq{CS},\eq{topdiv1}  
(Maxwell-Chern-Simons equations) in terms of the electric $\vec E$ and magnetic $\vec B$ fields:
\beq\label{MCS1}
\vec{\nabla}\times \vec{B} - \frac{\partial \vec{E}}{\partial t} = \vec J + c \left(\dot{\theta} \vec{B} - \vec{P} \times \vec{E}\right), 
\eeq
\beq\label{MCS2}
\vec{\nabla}\cdot \vec{E} = \rho + c \vec{P} \cdot \vec{B},
\eeq
where $(\rho, \vec J)$ are the electric charge and current densities; the second pair of Maxwell equations is not modified. 
As we will now argue, the Maxwell-Chern-Simons theory describes $P$ and $CP$ odd effects in hot QCD matter; for a more detailed discussion see \cite{Kharzeev:2009ry}.

\section{Chiral magnetic effect}

\subsection{Charge separation}

Consider now a configuration  where an external magnetic field $\vec B$ pierces a domain with $\theta \neq 0$ inside;  outside $\theta=0$. Let us assume first that the field $\theta$ is static, $\dot\theta = 0$. Assuming that the field $\vec B$ is perpendicular to the domain wall, 
we find from \eq{MCS2} that the upper domain wall acquires the charge density per unit area $S$ of  \cite{Kharzeev:2007tn}
$
(Q/S)_{up}  = +\ c\ \theta B
$
while the lower domain wall acquires the same in magnitude but opposite in sign charge density
$
(Q/S)_{down}  = -\ c\ \theta B
$.
Static electric dipole moment is a signature of ${P}$, ${T}$ and ${CP}$ violation (we assume that ${CPT}$ invariance holds). The spatial separation of charge will induce the corresponding electric field $\vec E = c\ \theta\ \vec B$. The mixing of pseudo-vector magnetic field $\vec B$ and the vector electric field $\vec E$ signals violation of ${P}$, ${T}$ and ${CP}$ invariances.

If the domain is due to the fluctuation 
of topological charge in QCD vacuum, its size is on the order of QCD scale, $L \sim \Lambda_{\rm QCD}^{-1}$, $S \sim \Lambda_{\rm QCD}^{-2}$. This means that to observe an electric dipole moment
in experiment we need an extremely strong magnetic field $eB \sim   \Lambda_{\rm QCD}^{2}$. Fortunately, such fields exist during the early moments of a relativistic heavy ion collision \cite{Kharzeev:2007jp,Skokov:2009qp}. Here we have assumed that the domain is static; this approximation requires the characteristic time of topological charge fluctuation $\tau \sim 1/\dot{\theta}$ be large on the time scale at which the magnetic field $B$ varies. This assumption is only marginally satisfied in heavy ion collisions, and so we now need to consider also the case of $\dot\theta \neq 0$.
  
\subsection{The chiral induction}
 
 Consider now the domain where $| \vec{P} | \ll \dot{\theta}$, i.e. the spatial dependence of $\theta(t, \vec x)$ is much slower than the dependence on time \cite{Kharzeev:2007jp}. Again, we will expose the domain to an external magnetic field $\vec B$ with $\vec{\nabla}\times \vec{B} = 0$, and assume that no external electric field is present.  In this case we immediately get from \eq{MCS1} that there is an induced current \cite{Fukushima:2008xe}
 \beq\label{chimag}
 \vec{J} = - c\ M\ \vec{B} = - \frac{e^2}{2 \pi^2}\ \dot\theta\ \vec{B}.
 \eeq
Note that this current directed along the magnetic field $\vec B$ represents a ${P}-$, ${T}-$ and ${CP}-$odd phenomenon and of course is absent in the "ordinary" Maxwell equations. Integrating the current density over time (assuming that the field $\vec B$ is static) we find that when $\theta$ changes from zero to some $\theta \neq 0$, this results in a separation of charge and the electric dipole moment.

\subsection{Charge separation at finite baryon density}
 
 At finite baryon density, the charge separation and the chiral induction can occur even without a magnetic field if the angular momentum is present \cite{Kharzeev:2007tn,Kharzeev:2004ey}.  Indeed, let us introduce a matter velocity field $V_{\mu}$; then at finite baryon density $\mu$ Eq. \eq{CS} acquires the following additional term: 
 \beq 
\label{1}
{\cal L}_B = 
- N_c \frac{e \mu}{4 \pi^2}\cdot    \eps\dm \theta\ (\d_{\lambda}V_{\nu} ) A_{\sigma}.
 \eeq
This term has been used in the studies of the axial current in cold dense quark matter \cite{SZ,MZ}; however it is easy to see that it also 
 induces the electric charge density on the topological domain walls  \cite{Kharzeev:2007tn}:
\beq
\label{2}
\rho_B =\frac{\delta  L_B}{\delta A_0}=
N_c \frac{e \mu }{4 \pi^2}\cdot    \epsilon^{ijk}\d_i \theta\ (\d_{j}V_{k} ) =
  N_c \frac{e \mu}{2 \pi^2}\cdot      \left(\vec{\nabla}\theta\cdot \vec{\Omega}\right),
 \eeq
where $ \vec{\Omega}$ is the angular velocity of the rotating system, $ 2\epsilon_{ijk} \Omega_k= (\d_i V_j-  \d_{j}V_{i} )$. The result (\ref{2})  shows that there is a charge separation along the direction of $ \vec{\Omega}$. In a recent paper \cite{Son:2009tf} it has been shown that the chiral anomaly affects also the relativistic hydrodynamics, in particular by giving rise to the charge separation phenomenon in a rotating fluid. 
 
 \section{Experimental observables}

 Since there is no {\it global} violation of {\cal P} and {\cal CP} invariances in QCD, the sign of the expected charge asymmetry should fluctuate from event to event. 
The experimental variable sensitive to this effect has been proposed by Voloshin \cite{Voloshin:2004vk}, and the first preliminary results have been 
reported in \cite{Selyuzhenkov:2005xa,Voloshin:2008jx}. At this Conference, the STAR Collaboration has presented a refined and extended analysis resulting in an evidence for the $P$-odd charge separation effect \cite{Voloshin:2009hr}. Numerous mundane backgrounds have been examined, and none of them could so far explain the observed effect \cite{Voloshin:2009hr}. It is clear that a dedicated experimental program of studying topological effects in hot QCD matter is necessary to understand fully this intriguing observation -- for example, the energy and the mass number dependencies, as well as the measurement of charge asymmetries for identified particles would allow to test the proposed mechanism of local parity violation. 

The mechanism discussed in this talk requires a sufficiently large energy density for the sphaleron transitions to turn on, and for the quarks to separate by a distance comparable to the system size -- therefore, there has to be a deconfined phase. In addition the system has to be in a chirally symmetric phase -- indeed, in a chirally broken phase, the chiralities of quarks could flip easily causing dissipation of the induced current.  The experimental and theoretical studies of parity--odd charge asymmetries in heavy ion collisions can significantly improve our understanding of the topological structure of QCD, and help to detect the creation of a deconfined and chirally symmetric phase of QCD matter.

  \section*{Acknowledgments}
I am grateful to K. Fukushima,  L. McLerran,  H. Warringa and A. Zhitnitsky for enjoyable and productive collaborations.
This work was supported by the U.S. Department of Energy under Contract No. DE-AC02-98CH10886.

\end{document}